\documentclass[12pt]{iopart}

\usepackage[pdftex]{graphicx}
\usepackage{iopams}
\usepackage{epstopdf}
\usepackage{cite}
\usepackage{gensymb}

\begin{document}
\def\be{\begin{equation}}
\def\ee{\end{equation}}
\def\bea{\begin{eqnarray}}
\def\eea{\end{eqnarray}}
\def\rp{r_{+}}
\def\rmm{r_{-}}

\title{Potts Models with Invisible States on General Bethe Lattices}

\date{March 2013}
\author{N. Ananikian and N.Sh. Izmailyan}
\address{A.I. Alikhanyan National Science Laboratory,
Alikhanian Br.2, 375036 Yerevan, Armenia.\\{\it and}\\
 Applied Mathematics Research Centre, Coventry University, Coventry CV1 5FB, England.}

\author{D.A. Johnston}
\address{Dept. of Mathematics and the Maxwell Institute for Mathematical
Sciences, Heriot-Watt University,
Riccarton, Edinburgh, EH14 4AS, Scotland}

\author{R. Kenna}
\address{Applied Mathematics Research Centre, Coventry University, Coventry, CV1 5FB, England}

\author{R.P.K.C.M. Ranasinghe}
\address{Department of Mathematics, University of Sri Jayewardenepura,
Gangodawila, Sri Lanka.}


\begin{abstract}

The number of so-called invisible states which need to be added to the $q$-state Potts model to transmute its phase transition from continuous to first order has attracted recent attention.
In the $q=2$ case, a Bragg-Williams, mean-field approach necessitates four such invisible states while a $3$-regular, random-graph formalism requires seventeen. 
In both of these cases, the changeover from second- to first-order behaviour induced by the invisible states  is identified through the tricritical point of an equivalent Blume-Emery-Griffiths  model.

Here we investigate the generalised Potts model on a Bethe lattice with $z$ neighbours.
We show that, in the $q=2$ case, $\displaystyle{r_c(z) =  {4 z \over 3(z-1)} \left( {z-1 \over z-2}\right)^z}$ invisible states are required to manifest the equivalent Blume-Emery-Griffiths  tricriticality.
When $z=3$, the $3$-regular, random-graph result is recovered, while
$z \rightarrow \infty$ delivers the Bragg-Williams, mean-field result.

\end{abstract} 

\maketitle


\section{Introduction}
The ferromagnetic $q$-state Potts model is defined through the Hamiltonian  
\begin{equation}
 {\cal H}_q =   -   \sum_{\langle ij \rangle}  \delta_{\sigma_i, \sigma_j} \; ,
\label{P1}
\end{equation}
with nearest-neighbour interactions between spins $\sigma_i$, defined at the sites $i$ of a suitable $d$-dimensional lattice \cite{Potts}.
In the standard set-up, the spins $\sigma_i$ each take one of $q$ possible values, sometimes referred to as ``colours''.
A phase transition is induced by breaking the underlying $q$-fold symmetry of the model and the nature of the transition, including its order, is a function of $q$. 
In $d=2$ dimensions, the Potts model has a second-order phase transition for $q \le 4$ and a first-order transition for higher $q$-values. 
For $d\ge 3$ dimensions, only the 2-state Potts (Ising) model has a continuous transition and transitions for higher $q$-values are of first order. 

Recent experimental studies have suggested that some models with $q$-fold symmetry breaking in two dimensions do  not display the same order of transition as the corresponding ferromagnetic  $q$-state Potts model \cite{q3,q31,q32}.
Motivated by such discrepancies, Tamura {\emph{et al.}} investigated an extended Potts model with a number of ``colourless'' or ``invisible''   states \cite{TT1,TT2,TT3,TT4}. 
These redundant states do not contribute to the internal energy of the system, nor do they alter  its symmetry
or the number of ground states available.
However, they  change the entropy of the model as they increase the overall number of microstates available to the system.
Tamura {\emph{et al.}} showed that such invisible states can change the order of a phase transition. 

The new Hamiltonian introduced in Refs.\cite{TT1,TT2,TT3,TT4} is
\begin{equation}
 \label{eq:original_Hamiltonian}
  {\cal H}_{(q,r)} = -  \sum_{\langle i,j \rangle}
  \delta_{s_i, s_j} 
  \sum_{\alpha = 1}^{q} \delta_{s_i, \alpha} \delta_{s_j, \alpha} \; ,
  \,\,\,\,\,\,\,
  s_i = 1, \cdots, q, q + 1, \cdots, q + r,
\end{equation}
where the second summation ensures that only the first $q$ spins contribute to the Hamiltonian (i.e. to the energy). 
While the remaining $r$ spins do not contribute to the Hamiltonian, they are  traced over in the partition function and thus contribute to the entropy. 
The new Hamiltonian defines a ($q,r$)-state Potts model with $q$ visible and $r$ invisible states.

Using numerical simulations, Tamura {\emph{et al.}} found that the introduction of a sufficiently large number of invisible states changes the nature of the two-dimensional, $q \le 4$,  Potts-model phase transition  from (continuous) second order  to first order.
The strength of these first-order transitions increases with the addition of yet more invisible states; the latent heat increases and the transition temperature decreases.
On the analytic side, Tamura {\it et al.} also applied a Bragg-Williams, mean-field approximation to the ($q,r$)-state Potts models. For $q=2$, this delivers a second-order transition for $r = 1,\, 2$ and $3$ and a first-order transition for  $r \ge 4$. If $q \ge 3$, mean field theory gives a first order transition even in the ordinary Potts model and this transition remains first order with the introduction of invisible states.

The existence of a first-order transition to a low-temperature, broken-symmetry phase  for the ($q,r$)-state models has been proven rigorously using random-cluster methods in Ref.\cite{AvE} for $q>1$ and sufficiently large $r$. 
The transmutation of some second-order transitions into first-order transition by invisible states is therefore well established.  
In Ref.\cite{DesMalmini} the case of $q=2$ visible states was investigated  using 3-regular random graphs as an  alternative route to mean-field calculations. 
A curious feature was that seventeen invisible states were required to induce a first-order transition using this route, compared to that of Bragg-Williams which required only $r=4$ to effect this change.

Here we investigate the $q=2$, generalised Potts model defined on the Bethe lattice with  a general number of nearest neigbours $z$. In the absence of invisible states this model exhibits a continuous phase transition.
We derive a general formula for the critical number of invisible states $r_c(z)$ above which the transition transmutes to first order.  In the case $z=3$ this recovers the random-graph result of Ref.\cite{DesMalmini}.
In the  $z \rightarrow \infty$ limit, our formula recovers the Bragg-Williams mean-field result that $r=4$ invisible states are required to render the transition first order.

Following Ref.\cite{TT1} it is convenient to rewrite  equ.~(\ref{eq:original_Hamiltonian}) 
by introducing spins $\sigma_i$, where $\sigma_i = s_i$ if $s_i =1, \cdots, q$ and $\sigma_i=0$ otherwise.
This leads to an effective Hamiltonian ${\cal H}'_{(q,r)}$ with $q$ standard spins and one additional spin which does not contribute to the nearest neigbour energy term but is coupled to a temperature dependent external field,
 \begin{eqnarray}
 \label{0Potts}
 {\cal H}'_{(q,r)} =  -  \sum_{\langle i,j \rangle} \delta_{\sigma_i, \sigma_j}
\sum_{\alpha = 1}^q \delta_{\sigma_i, \alpha} \delta_{\sigma_j, \alpha}
 - T \ln r \sum_i \delta_{\sigma_i, 0} \; ,
 \,\,\,\,\,\,\,
 \sigma_i = 0,1,\cdots,q.
\end{eqnarray} 
By construction, the partition functions for ${\cal H}_{(q,r)}$ and ${\cal H}'_{(q,r)}$ are identical, so we may employ whichever formulation is most convenient. 
For the particular case where $q=2$, direct consideration of the Boltzmann  weights  in the latter formulation shows equivalence to a Blume-Emery-Griffiths (BEG)  Hamiltonian,
\begin{equation}
\label{BEG}
 {\cal H}_{\rm BEG} = - {1\over 2} \sum_{\left\langle i,j \right\rangle}
  t_i t_j     - {1\over 2} \sum_{\left\langle i,j \right\rangle}    t_i^2 t_j^2 
 - \mu \sum_i \left( 1 - t_i^2 \right), \quad
  t_i = +1,\, 0,\, -1
\end{equation}
with a temperature-dependent, crystal-field term $\mu = T \ln r$ and equal couplings for the  two nearest-neighbour interaction terms \cite{BEG}.

\begin{figure}[t]
\begin{center}
\includegraphics[height=5cm]{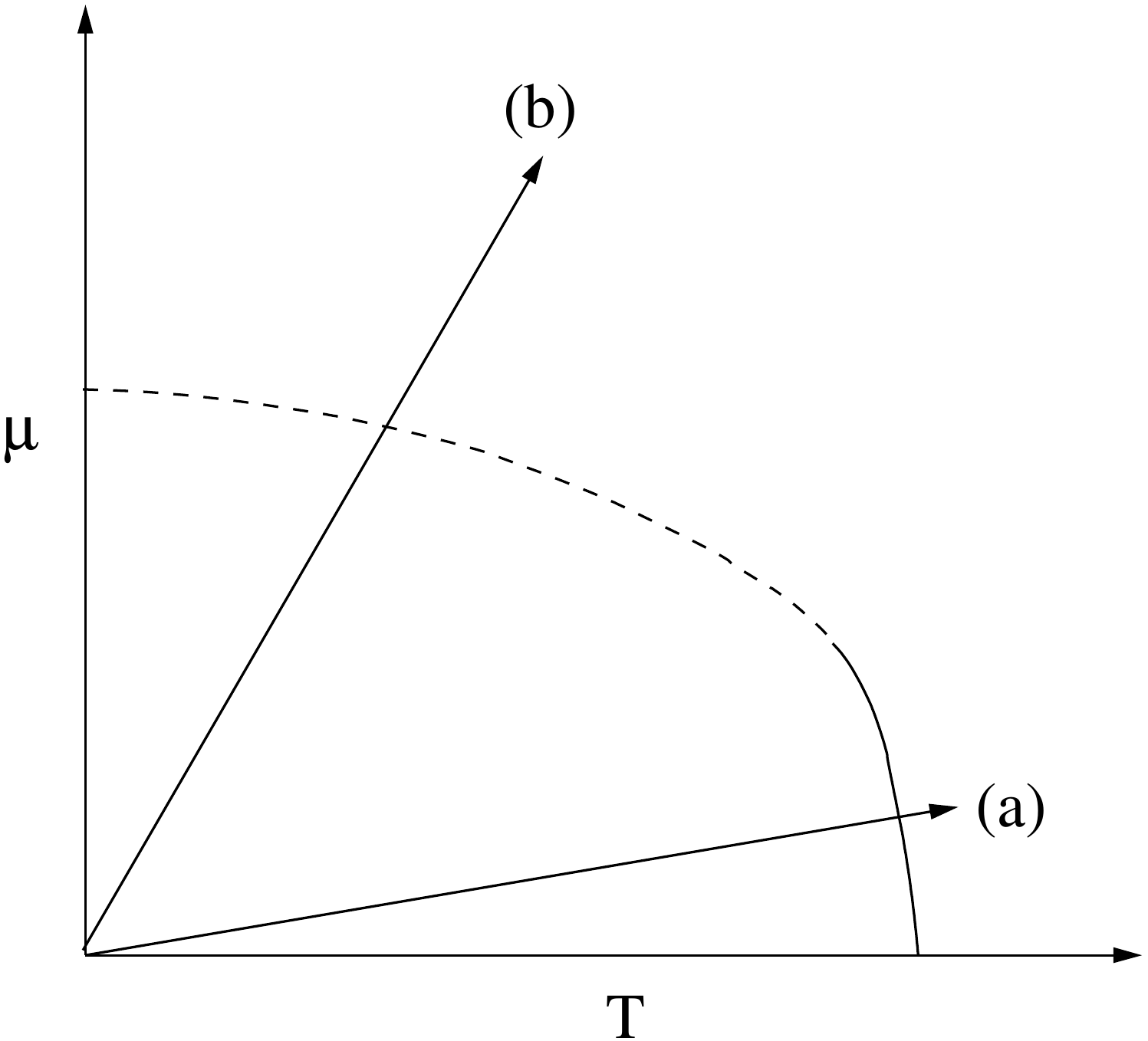}
\label{BEGgraph} 
\caption{A schematic drawing of an equal-coupling, mean-field, BEG-model phase diagram
in the $\mu,\, T$ plane. The second-order transition regime is shown as a solid line and the first-order region  as a dashed line. The arrowed, sloped lines marked (a) and (b) represent the trajectory of the system as $T$ is increased for different values of $r$, using $\mu = T \ln r$. }
\end{center}
\end{figure}
The tactic is then to exploit knowledge of the  phase diagram of the BEG model to investigate the effects of varying the number of invisible states $r$ in the equivalent $(2,r)$-state Potts model. Analytical calculations are possible in various circumstances.   
In particular, Tamura {\it et al.} used a Bragg-Williams mean-field calculation to show that four invisible states were sufficient to transmute the $(2,r)$-state, Potts-model transition into a first-order transition.
In Ref.\cite{DesMalmini}, on the other hand, it was found that a different mean-field calculation on 3-regular, random graphs required seventeen invisible states to effect such a change. 
The mechanism for changing the order of the transition in both cases is identical and can  be understood  by examining the phase diagram of the BEG model in the $\mu, T$ plane (Fig.\ref{BEGgraph}).

The BEG model manifests a line of phase transitions which changes from first to second order at a tri-critical point.
The $(2,r)$-state Potts system accesses the BEG phase transitions by following lines of increasing slope for increasing $r$ as $T$ is varied. 
For $r=1$ the  $(2,r)$-state Potts model is identical to the BEG model with vanishing crystal field. 
For small values of $r$, the generalised Potts model remains in this universality class.
However, for sufficiently large values of $r$, such a line traverses the first-order portion of the BEG phase diagram, rather than the second-order part.  
The number of invisible states required to transmute the phase transition from second to first order is therefore given by the position of the tri-critical point on the BEG-model phase diagram.

In the next section, we follow Ref.\cite{Ananikian1,Ananikian2} and use recursion relations to derive the tricritical point of the BEG model on the Bethe lattice with equal couplings for the two nearest-neighbour interaction terms. 
This  is compared to a saddle-point calculation on regular random graphs in Section~3. Although the methods used are rather different the solutions  are identical. 
It proves easier to generalize the Bethe-lattice calculation to an arbitrary number of nearest neighbours, and we give  the general formula for the tri-critical point, and hence the critical number of invisible states, for any number $z$ of nearest neighbours on the Bethe lattice.

\section{The BEG model on a Bethe lattice and its tri-critical point}

\begin{figure}[t]
\begin{center}
\includegraphics[height=5cm]{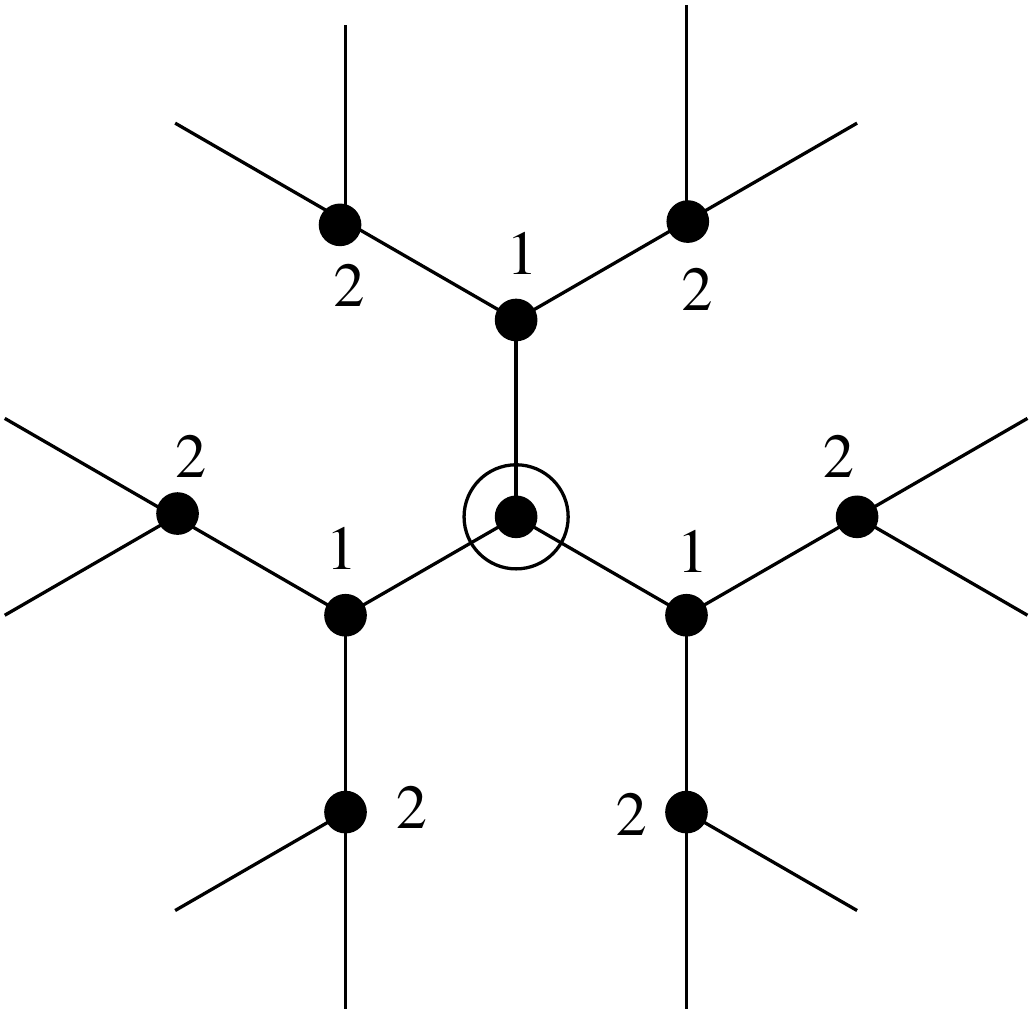}
\label{Bethepic} 
\caption{Two shells around  the central vertex (circled) of a Bethe lattice with $z=3$ nearest neighbours. In the full lattice the branching continues ad infinitum. For the recursive calculation of the partition function  the spin $t_0$ resides on the central vertex.}
\end{center}
\end{figure} 
The Bethe lattice offers a convenient way to formulate mean-field models in statistical mechanics since the hyperbolic nature of its geometry means that it is effectively infinite dimensional \cite{Bethe}.  
The shell-like nature of its construction also means that statistical mechanical models formulated on the Bethe lattice lend themselves to exact solutions via recursion relations. 
The first two generations, or shells, of a Bethe lattice with $z=3$ neighbours are shown in Fig.(\ref{Bethepic}).
The statistical mechanical behaviour of spin models defined on the Bethe lattice is calculated ``deep within'' the lattice, disregarding the effect of the boundary, which contains as many points as lie within the lattice itself. 
Regular random graphs offer an alternative way to perform what is effectively the same calculation since they appear locally identical to the Bethe lattice, but with branches that are closed off by generically large loops.  

Our aim is to evaluate the partition function for the BEG model on a Bethe lattice and, in particular, to determine the tri-critical point which will allow us to find the critical number of invisible states required to effect a first-order transition. 
This may be done in a standard manner for the Bethe lattice by evaluating the partition function recursively, shell by shell, starting at the central vertex given by $i=0$ \cite{Ananikian1,Ananikian2,CT,CM}. 
The partition function can be written as
\begin{equation}
Z = \sum_{\{t\}} \exp \left(   -\beta   {\cal H}_{\rm BEG}  \right),
\end{equation}
or, more explicitly, as
\begin{equation}
Z = \sum_{\{t\}} \exp \left(  {\beta \over 2} \sum_{\left\langle i,j \right\rangle}
  t_i t_j     + {\beta \over 2} \sum_{\left\langle i,j \right\rangle}    t_i^2 t_j^2 
 -  \ln{r} \sum_i t_i^2   \right),
\end{equation}
where we have used $\beta \mu = \Delta=\ln r$ and ignored an inessential constant. 
This may be separated into the contribution of the central spin, $t_0$, and the $z$ branches emerging from it
\begin{equation}
Z = \sum_{t_0} r^{- t_0^2}  \left[ g_l (t_0)\right]^z,
\end{equation}
where the branch partition function $g_l(t_0)$ with $l$ shells is given by
\begin{eqnarray}
g_l(t_0) &=& \sum_{t \ne t_0}  \exp \left( {\beta \over 2} t_0 t_1 + {\beta \over 2} t_0^2 t_1^2 +  {\beta \over 2} \sum_{\left\langle i,j \right\rangle}
  t_i t_j     + {\beta \over 2} \sum_{\left\langle i,j \right\rangle}    t_i^2 t_j^2  \right)
  \times r^{ -  \sum_i t_i^2}   \nonumber \\
&{}& 
\end{eqnarray}
This  in turn may be written recursively as
\begin{equation}
g_l(t_0) = \sum_{t_1} \exp \left( {\beta \over 2} t_0 t_1 + {\beta \over 2} t_0^2 t_1^2   \right) 
r^{-  t_1^2 }
\left[ g_{l-1} (t_1) \right]^{z-1} .
 \end{equation}
Defining the  ratios 
\begin{equation}
x_l = { g_l(-1) \over g_l(0)} \quad \quad {\mbox{and}} \quad \quad 
y_l = { g_l(+1) \over g_l(0)} 
\label{recursiong}
\end{equation}
allows the branch partition function recursion relations to be recast as 
\begin{equation}
\label{recursion}
x_l = {r + x_{l-1}^{z-1} e^{\beta} + y_{l-1}^{z-1} \over r + x_{l-1}^{z-1}  + y_{l-1}^{z-1}  } , \quad \quad \quad
y_l = { r + x_{l-1}^{z-1} + y_{l-1}^{z-1}  e^{\beta} \over r + x_{l-1}^{z-1}  + y_{l-1}^{z-1}  } .
\end{equation}
The different phases of the BEG model appear as different fixed points of these recursion relations as the  parameters $r$ and $\beta$ are varied ($x_l = x_{l-1} \equiv x$ and $y_l = y_{l-1} \equiv y$). 
Following Ref.\cite{Ananikian1,Ananikian2} we define
\begin{equation}
u = {1 \over 2} ( x + y - 2), \quad \quad v= {1 \over 2} ( x - y) ,   \quad \quad
b = {e^{\beta}-1 \over 2} \; ,
\end{equation}
and rewrite the fixed point equations  as
\begin{eqnarray}
r^2 &=& { 4 ( b - u)^2 \over u^2 - v^2} \left[ ( u+1)^2 - v^2 \right]^{z-1}\\
1 &=& { u - v \over u+v} \left( { u  + v + 1 \over u  - v + 1} \right)^{z-1} \;  .  \nonumber
\end{eqnarray}
As discussed in Ref.\cite{Ananikian1} there are two  families of solutions  to the recursion relations.
The first is given by 
\begin{equation}
v = 0, \quad \quad \quad 
r = {2 ( b - u) ( u+1)^{z-1} \over u} \; .
\end{equation}
The second solution has 
\begin{equation}
r^2 = { 4 ( b - u)^2 \over u^2 - v^2} \left[ ( u+1)^2 - v^2 \right]^{z-1} 
\label{star}
\end{equation}
 with
\begin{equation}
1 -  { (z-1)u \over u+1}  + \left( { v \over u+1} \right)^2 {\cal{F}(}u,v) = 0 \; ,
\end{equation}
where, for odd $z$
\begin{eqnarray}
 {\cal{F}}(u,v)  =  \left( C^{z-1}_2 -  {C^{z-1}_3 u \over u+1} \right)+ \ldots + \left({ v \over u+1 } \right)^{z-3} , 
\end{eqnarray}
and for even $z$
\begin{eqnarray}
 {\cal{F}}(u,v) =               \left( C^{z-1}_2 - {C^{z-1}_3 u \over u+1} \right) + \ldots 
+ \left( z-1 - {u \over u+1} \right) \left( {v \over u+1 } \right)^{z-4} 
\end{eqnarray}
where the $C^{z-1}_n$  are binomial coefficients of an expansion in $v/(u+1)$.

The $\lambda$-line of critical points in the BEG model is determined by the equality of the two solution sets 
which occurs when 
\begin{equation}
r = { 2 ( b - u) \over u} ( u + 1 )^{z-1} \; , \quad \quad 
u = \frac{1}{z-2} \; .
\label{18}
\end{equation}
The tricritical point on this line is given by the solution of 
\begin{equation}
{\delta r^2 \over \delta u} = {\partial r^2 \over \partial u} + {\partial r^2 \over \partial v^2} { \partial v^2 \over \partial u} =0 \; ,
\end{equation}
where $r(u,v)$ is given in equ.(\ref{star}). This gives
\begin{eqnarray}
{ u + 1 \over b - u} = z-2 + { z-3 \over 2 z} { 1 \over u}  \; .
\label{20}
\end{eqnarray}
From equs.(\ref{18}) and (\ref{20}), we obtain values of $b$ and $r$ at the tri-critical point on  a Bethe lattice with $z$ neighbours as
\begin{equation}
b_c(z) = { 5 z - 6 \over 3 ( z-2)^2 }
\end{equation} 
and 
\begin{equation}
r_c(z) = { 4 z \over 3 ( z-1)} \left( z-1 \over z-2 \right)^{z}  .
\end{equation}
Referring back to the equivalence between BEG model and the ($2,r$)-state Potts model we see that the Potts model 
displays a first-order transition for $r>r_c$ invisible states.

Since the result applies for general $z$ we have, for instance,
\begin{eqnarray}
z &=& 3, \; \; \; r_c(3) = 16 \; , \nonumber \\
z &=& 4, \; \; \; r_c(4) = 9 
\end{eqnarray}
and we can also take the $z \rightarrow \infty$ limit to find $r_c( \infty) = 4 e / 3 \simeq 3.624$. 
The monotonicity of $r_c(z)$ as a function of $z$ is shown   in Fig.(3).
\begin{figure}[t]
\begin{center}
\includegraphics[height=7cm]{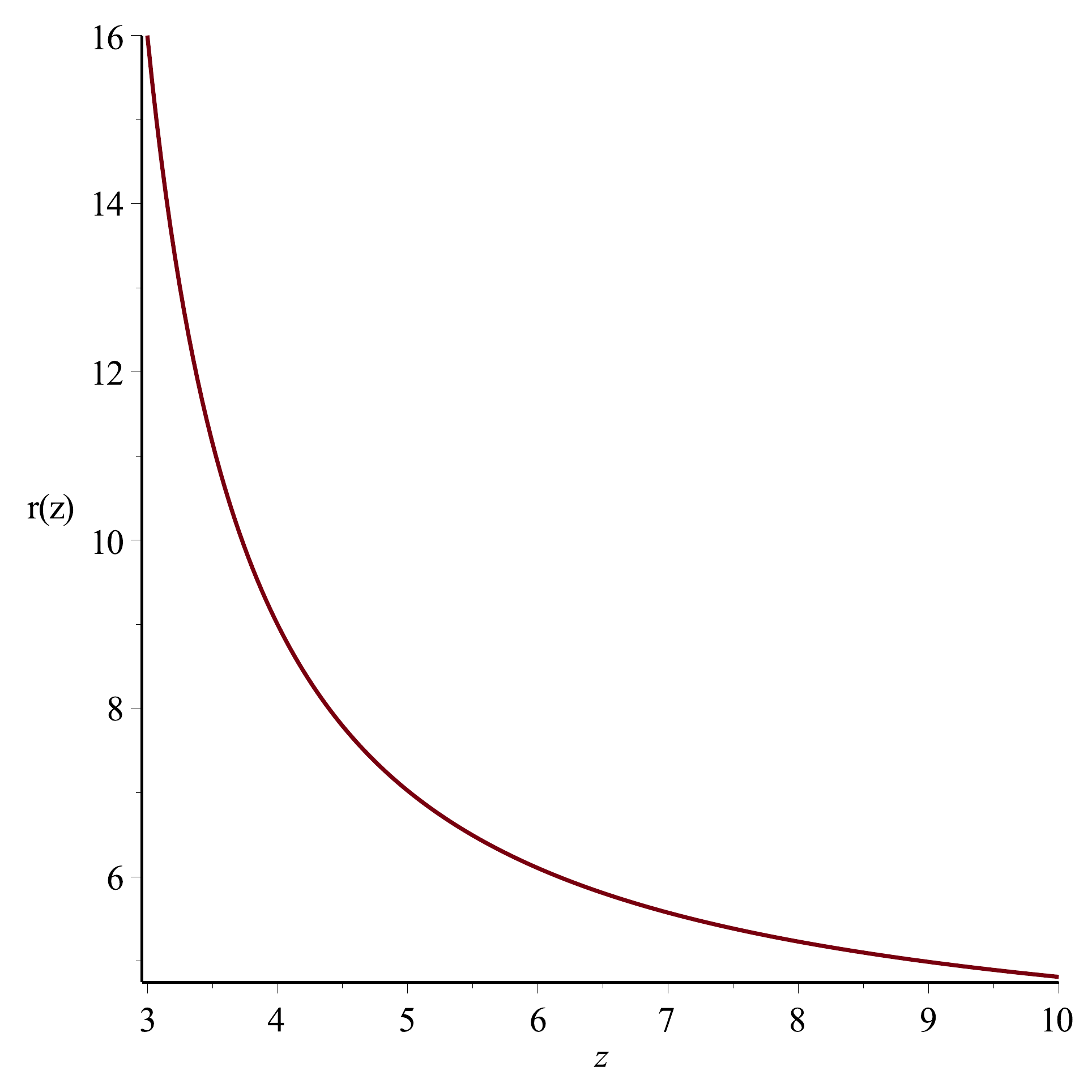}
\label{rc} 
\caption{The critical number of invisible states $r_c(z)$ above which the ($2,r$) state Potts model displays a first order transition, plotted against the number of neighbours $z$ in the Bethe lattice.}
\end{center}
\end{figure}

\section{Comparison with the BEG model on random graphs}

Another approach to mean field theory in statistical mechanics is to consider the models on regular random graphs.
As we have noted, they are clearly related to the same models on the Bethe lattice since the local environment for the spins is identical. The (generically) large loops which close the branches of the Bethe lattice to give the corresponding regular random graph turn out not to affect the critical behaviour, at least for the ferromagnetic transitions considered here \cite{JP,JP2}.   

It is possible to enumerate undecorated 3-regular random graphs by considering them to be generated by the ``Feynman diagram'' expansion of a scalar integral, rather than a path integral as in a quantum field theory or an integral over matrices as in a matrix model \cite{0,00}. 
The number of 3-regular random graphs with $n$ vertices is given by evaluating the integral
\begin{equation}
N_n = {1 \over 2 \pi i} \oint { d \lambda \over
\lambda^{2n + 1}} \int_{-\infty}^{\infty} d \phi \; \exp \left( -\frac{1}{2} \phi^2 + \frac{\lambda}{3} \phi^3 \right)
\end{equation}
using a perturbative expansion of  the $\phi^3$-term to the required order. 
Other families of random graphs can be enumerated in a similar fashion by simply replacing the potential, for instance by using  a $\phi^r$ term for $r$-regular random graphs.
The graphs may be decorated with the appropriate weights, both edge and vertex, for the statistical mechanical model under consideration by evaluating a similar integral for an ``action'' which generates the correct weights when expanded perturbatively. In the case of the BEG model on 3-regular random graphs such an integral is given by    
\begin{equation}
\label{ZnIsing}
Z_n(\beta) \times N_n = {1 \over 2 \pi i} \oint { d \lambda \over
\lambda^{2n + 1}} \int {d \phi_1 d \phi_2 d \phi_3 \over 2 \pi \sqrt{\det K}}
\; \exp (- S_{\rm{BEG}} ) \; ,
\end{equation}
where the BEG action is
$S_{\rm BEG}$ 
\begin{eqnarray}
S_{\rm BEG} &=& { 1 \over 2 } ( \phi_1^2 +\phi_2^2 + \phi_3^2)  - a ( \phi_1 \phi_3 + \phi_2 \phi_3) -  { \lambda \over 3 } ( \phi_1^3 + \phi_2^3 + \sigma  \,  \phi_3^3) 
\label{SBEGeq}
\end{eqnarray}
and $K$ is the propagator evaluated from the inverse of the quadratic coefficients.
The coefficients in $S_{\rm BEG}$ are related to the Hamiltonian couplings by
\begin{eqnarray}
\label{BEGrels2}
e^{- \beta} &=& {a^2 \over 1 - a^2} \; , \nonumber \\
\sigma &=& a^3 \, e^{\beta \mu} = a^3 r \; ,  
\end{eqnarray}
so the physical range of the coefficient
\begin{equation}
a = \sqrt{{ 1 \over e^{\beta} + 1}} \nonumber
\end{equation} 
in $S_{\rm BEG}$ is $0<a<1/\sqrt{2}$. The vertex coupling $\lambda$ may be scaled out of $S_{\rm BEG}$ and the leading contribution in the thermodynamic limit $n \rightarrow \infty$ evaluated using the saddle point equations
\begin{eqnarray}
\label{saddle}
{\partial S_{\rm BEG} \over \partial \phi_1} &=& \phi_1 -a \phi_3 - \phi_1^2   = 0 \; ,\nonumber \\
{\partial S_{\rm BEG} \over \partial \phi_2} &=& \phi_2 -a \phi_3 - \phi_2^2    =0 \; ,\\
{\partial S_{\rm BEG} \over \partial \phi_3} &=&  \phi_3 -a (\phi_1 + \phi_2)  - a^3 r \phi_3^2  =0 \nonumber 
\end{eqnarray}
whose various solutions then delineate the phase diagram.

Consideration of other spin models on such regular random graphs has shown that the content of the saddle point equations is identical to the fixed point equations obtained when the models are formulated on the corresponding Bethe lattice with the same number of neighbours. Although it is not immediately apparent that this is also the case for the BEG recursion relations in equ.(\ref{recursion}) and the saddle point equations in equ.(\ref{saddle}), we can show this is so by using the first two equations of (\ref{saddle}) to write the third as
\begin{equation}
\label{phitilde}
\tilde{\phi}_3 = {a^2 \over 1 - 2a^2} \; ( \phi_1^2 + \phi_2^2 ) + {a^2 r \over 1 -2 a^2} \;  \tilde{\phi}_3^2 \; ,
\end{equation}
where we have rescaled $\phi_3 \rightarrow \tilde{\phi}_3 / a$ for convenience.
The first two equations may then be rewritten using equ.(\ref{phitilde}) as
\begin{eqnarray}
\phi_1 &=& \tilde{\phi}_3 + \phi_1^2 = {1-a^2 \over 1 - 2a^2} \;  \phi_1^2  + {a^2 \over 1 - 2a^2} \;  \phi_2^2  + {a^2 r \over 1 -2 a^2} \;  \tilde{\phi}_3^2 \; ,\nonumber \\
\phi_2 &=& \tilde{\phi}_3 + \phi_2^2 = {a^2 \over 1 - 2a^2} \;  \phi_1^2   + {1 - a^2 \over 1 - 2a^2} \;  \phi_2^2  + {a^2 r \over 1 -2 a^2} \;  \tilde{\phi}_3^2
\end{eqnarray}
and taking the ratio of these with equ.(\ref{phitilde}) recovers the fixed point equations for the recursion relations in equ.(\ref{recursion}) for $z=3$ with $x = \phi_1 / \tilde{\phi}_3$ and $y = \phi_2 / \tilde{\phi}_3$.

Similar manipulations of the saddle point equations may be used to demonstrate the equivalence with the Bethe lattice fixed points for general $z$, so  conclusions drawn about the Bethe lattice phase diagram with $z$ neigbours may also be taken to apply to the model on $z$-regular random graphs.

\section{Discussion}
The number of invisible states required to obtain a first order-transition is determined by the {\it position} of the tri-critical point in the BEG model, so it is a non-universal quantity. It is therefore no surprise that it depends on the details of the lattice under consideration, such as the number of neighbours for the Bethe lattice considered here. We have seen, however, that the fixed point of Bethe lattice recursion relations and the saddle point equations which determine the phase diagram on the  regular random graphs have the same content.
The tri-critical point $r_c(z)$ on $z$-regular random graphs is thus determined by the same equations which give the fixed points of the recursion relations on the Bethe lattice with $z$ neighbours and the values of $r_c(z)$ are identical in these cases. The previous calculation in Ref.\cite{DesMalmini}  of $r_c(3)$ for 3-regular random graphs agrees with $r_c(3)=16$ found here for the Bethe lattice. 

It is also interesting to note that the result found for the Bragg-Williams approximation, where $4$ invisible states are sufficient to produce  a first-order transition with a ($2,r$) state Potts model, is consistent with the $z \rightarrow \infty$ value calculated here on the Bethe lattice. In this case the nearest neighbour environment with $z \rightarrow \infty$ is closer to that in standard mean field theory where we think of the model as living in a high dimensional space where $z$ is also  large. 

The fact that invisible states  can produce first-order transitions on Bethe lattices
is evidence for the generality of the phenomenon, while its dependency on the coordination number  $z$ indicates 
the non-universality of the mechanism. The explicit calculations here have focussed on the ($2,r$)-state Potts model  since, as we have noted,  mean field theory in all its variants gives a first order transition for the standard  $q \ge 3$ state Potts model. The happy coincidence of the correspondence with the BEG model then allows the explicit determination of $r_c(z)$. On {\it planar} random graphs continuous Potts transitions exist for $q=2,3,4$ and we have already remarked in \cite{DesMalmini} that the critical number of invisible states in the 
($2,r$) state Potts model on  $4$-regular, planar random graphs may be transcribed from the solution using matrix models \cite{3M1,3M2} of the BEG model on such graphs  
to give $r_c = 223$ states. In principle, the matrix models for the $q=3,4$ state Potts models with a suitable external field coupling to an invisible state in the manner of equ.~(\ref{0Potts}) would allow a similar determination for $q=3,4$ in a non-mean-field context.

\section{Acknowledgements}
N.A., N.Sh.I. and R.K. were partially supported by an FP7 EU IRSES project 
and N.Sh.I. and R.K. by a Marie Curie International Incoming Fellowship within the 7th European Community Framework Programme.
R.P.K.C.M.R. would like to thank the University of Sri Jayewardenepura for foreign leave and the Maxwell Institute for Mathematical
Sciences for hospitality during this work.


\end{document}